\documentclass[aps,twocolumn,superscriptaddress]{revtex4}
\usepackage{amsmath,amssymb}
\usepackage{graphics,graphicx}
\usepackage{dcolumn,bm}
\usepackage{psfrag}
\usepackage{color}
\usepackage{multirow}
\newcommand{\cu}
{\affiliation{Department of Physics, University of Calcutta,
92 Acharya Prafulla Chandra Road, Kolkata 700009, India.}}

\newcommand{\victo}
{\affiliation{Department of Physics, Victoria Institution (College),
78B Acharya Prafulla Chandra Road, Kolkata 700009, India.}}

\begin{document}

\title{Opinion formation models with extreme switches and disorder: critical behaviour and dynamics}

\author{Kathakali Biswas}
\victo
\cu
\author{Parongama Sen}
\cu
\begin{abstract}
In a three state kinetic exchange opinion formation model, the effect of extreme switches was considered in a recent paper. In the present work, we study  the same model with disorder. Here disorder implies 
that   negative interactions  may  occur with a probability $p$. In absence of  extreme switches,  the known critical point is at $p_c =1/4$ in the mean field model.   With a nonzero value of $q$ that denotes  the 
probability of such switches, the critical point is found to occur at $ p = \frac{1-q}{4}$ where 
the order parameter vanishes with a   universal value of the  exponent $\beta =1/2$.
Stability analysis 
of  initially ordered states near the phase boundary 
reveals the exponential growth/decay of the order parameter 
in the ordered/disordered phase with a   timescale diverging with exponent $1$.  The fully ordered state also 
relaxes  exponentially to its equilibrium value  with 
a similar behaviour of the associated  timescale.
Exactly at the critical points, the order parameter shows a   power law decay with time with exponent $1/2$. 
Although the critical behaviour remains mean field like, 
the system behaves more  like a two state model as $q \to 1$. At $q=1$ the model  behaves like a binary voter model with random flipping occurring with probability $p$.

\end{abstract}

\maketitle

\section{introduction}
	
To address the problem of opinion formation in a society \cite{soc_rmp,sen_chak,galam_book}, several models with three opinion states have been considered recently \cite{BCS,meanfield,nuno1,nuno2,sudip,vazquez,vazquez2,mobilia,lima,migu,cast,luca,hadz,gekle,galam3,sm_sb_ps,kb2}. Typically these opinions are taken as $\pm1$ and $0$, where $\pm1$ may represent extreme ideologies. 
In a recent paper \cite{kb2}, using a mean field kinetic exchange model, the present authors 
studied  the effect of  extreme switches of opinion, which is not usually  considered
in such models. 
Several interesting results were obtained, in particular, 
for the maximum probability of such a switch, the model  was shown to effectively reduce to a mean field Voter model beyond a transient time.
  In this paper we extend the previous work by including negative interaction between the agents which acts as a disorder. Such negative interactions have been incorporated in three 
state kinetic exchange models previously \cite{BCS,meanfield,nuno1,nuno2,sudip} and several properties have been studied in different dimensions.
However,  the effect of extreme switches and negative interaction both occurring together has not been studied earlier. Since 
these two features can occur simultaneously in reality, the dynamics of a model incorporating both
is worth studying. In absence of the extreme switches 
the critical point  
as well as the critical behaviour is known 
\cite{BCS,meanfield,nuno1}.  
The interest is primarily to see how the critical behaviour is   affected in presence of the extreme switches.

In the present two parameter model, representing the probabilities of negative interaction and 
extreme switches, in addition to obtaining the phase boundary and behaviour 
of the order parameter,  
we have  studied the dynamical behaviour close to the fixed point.  The relaxation of the 
 order parameter from a fully ordered state is also studied at and away from criticality. 
The static critical behaviour as well as the dynamical behaviour are found to be similar to the mean field  model  without extreme switches. 
However, we find that the nature of the phases in terms of  the densities of the three types of opinions is quite different. 
Especially, the case with maximum extreme switches in the presence of the negative interaction leads to
an interesting mapping to a disordered voter model. 
As a starting point, the mean field model has been studied where majority of the results can be obtained analytically. 
We derive the time derivatives  of the three densities of population in terms of the 
transition rates 
which are then either solved analytically or numerically. A small scale simulation is also made particularly to study  the finite size scaling behaviour of the order parameter. 

In section 2, the model is described. Results are presented  in section 3 and 
 and some further analysis are made in the last section which also includes the concluding remarks.

\section{The Model}

We have considered a  kinetic exchange model  for opinion formation  with three opinion values $0, \pm 1$.  Such states may represent the support for two candidates/parties and a neutral opinion \cite{kb2,US,kb1} or three different ideologies where $\pm 1$ represent radically different ones. 
The opinion of an  individual is updated by taking into account her  present opinion and an interaction with a randomly chosen 
 individual in the fully connected model. 
The time evolution of the opinion   of the $i$th  individual opinion denoted by 
$o_i(t)$, when she  interacts  with the $j$th individual, chosen randomly, is given by 
\begin{equation}
o_i(t+1)=o_i(t) + \mu  o_j(t),
\end{equation}
where $\mu$ is interpreted as an interaction parameter, chosen randomly. 
The opinions are bounded in the sense $|o_i| \leq 1$ at all times and therefore 
$o_i $ is taken as 1 (-1) if it is more  (less) than 1 (-1). There is no self-interaction so $i \neq j$ in general.  
The values of the interaction parameter are taken to be discrete: $\mu = \pm 1$ and $\pm 2$.
Here we take $|\mu| =1$ and $2$   with  probability  $1-q$ and $q$ respectively 
and  negative interactions, i.e., a negative value of $\mu$ occurs with probability $p$. 

\section{Results}

The mean field results are known for the limits $q=0$, any $p$  and also for $p=0$, any $q$. For  $q=0$, the system undergoes a order-disorder phase transition at $p = 1/4$.
For $p=0$ on the other hand, the fate of the  system starting from initially  ordered  configurations 
showed that it reached consensus for $q \neq 1$ while for $q=1$, there is a quasi conservation leading to a partially ordered state. Initially disordered states flow to a $q$ dependent frozen fixed point 
which is disordered for all $q$ \cite{kb2}. 
These results are ensemble averaged and valid in the thermodynamic limit. 

\subsection{Rate equations}

We first obtain the rate equations for the 
the densities of the three populations with opinion $0, \pm 1$,  
denoted by  $f_0, f_{\pm 1}$,  with $f_0 + f_{+1} + f_{-1} = 1$. The ensemble averaged order parameter
is  $\langle O \rangle = f_{+1} - f_{-1}$ with $-1 \leq \langle O \rangle \leq  1$. 

To set up the rate equations for the $f_i$'s, we need to treat the time variable as continuous. 
   Assume that the opinion changes from $i$ to $j$ ($i,j = 0,\pm1$)  
in time $\Delta t$  with the transition rate given by     $w_{i \rightarrow j}$.
Then we have the following set of $w_{ij}$'s:  
\begin{eqnarray}
w_{+1 \rightarrow +1} &= &f_{0}f_{+1} + f_{+1}^2(1-p) + f_{+1}f_{-1}p \nonumber \\
w_{0 \rightarrow +1}& = &(1-p)f_0f_{+1} + pf_0f_{-1}   \nonumber \\
w_{-1 \rightarrow +1}& = &q(1-p)f_{-1}f_{+1} + pqf_{-1}^2 \nonumber \\
w_{+1 \rightarrow 0}& = &(1-q)(1-p)f_{-1}f_{+1} + p(1-q)f_{+1}^2\nonumber \\
w_{0 \rightarrow 0}& = &f_0^2\nonumber \\
w_{-1 \rightarrow 0}& = &(1-q)(1-p)f_{-1}f_{+1} + (1-q)pf_{-1}^2\nonumber \\
w_{+1 \rightarrow -1}& =& q(1-p)f_{+1}f_{-1} + qpf_{+1}^2\nonumber \\
w_{0 \rightarrow -1}& =& (1-p)f_0f_{-1} + pf_{+1}f_0\nonumber \\
w_{-1 \rightarrow -1}& = &f_0f_{-1} + (1-p)f_{-1}^2 + pf_{-1}f_{+1}\nonumber
\end{eqnarray}

In general, we have $f_i(t+\Delta t) = f_i(t) + \sum_j w_{j\to i} \Delta t - \sum_j w_{i \to j}\Delta t$ 
such that taking $\Delta t \to 0$, we get   
\begin{widetext}
\begin{equation}
\frac{df_{+1}}{dt}  
=  qpf_{-1}^2 + f_0( (1-p)f_{+1} + pf_{-1} ) - (1-q)(1-p)f_{-1}f_{+1} - pf_{+1}^2,
\label{f+equation}
\end{equation}

\begin{equation}
\frac{df_{-1}}{dt} 
= qpf_{+1}^2 + f_0( (1-p)f_{-1} + pf_{+1} ) - (1-q)(1-p)f_{-1}f_{+1} - pf_{-1}^2.
\label{f-equation}
\end{equation}
 
\end{widetext}



\subsection{Steady states and critical behaviour}

Solving  equations  \ref{f+equation} and \ref{f-equation}, 
it is possible to obtain the time evolution of 
 the  order parameter $\langle O\rangle$ which satisfies
\begin{equation}
\frac{d\langle O\rangle}{dt} = [-qp - p + f_0(1 - p + qp)]\langle O\rangle.
\label{Oequation}
\end{equation}
To reach a steady state the right hand side of the above equation should be zero at $t \rightarrow \infty$.  It is obvious  that any initially disordered configuration will remain disordered. 

It was already observed in \cite{kb2} that the $q=1$ case is unique. Here also, it should be discussed separately. Precisely, for an ordered state to exist, 
$f_0 = 2p$ when $q=1$. However, we note that $f_0$ is expected to vanish very fast as there is no flux to the zero state from other states for $q=1$. Assuming $f_0$ vanishes within a transient 
time, one can  show  directly from the dynamical equations for the individual densities   that for an ordered state to exist, $p$ can only take the zero value when $q=1$.

Taking  the   origin  of time  as that when $f_0$ becomes  zero  and for $q=1$,   one can rewrite  equations \ref{f+equation} and \ref{f-equation} as
\begin{equation}
\frac{df_{+1}}{dt}  
=  pf_{-1}  - p f_{+1},
\label{f+equation_new}
\end{equation}
and
\begin{equation}
\frac{df_{-1}}{dt} 
=   pf_{+1} -p f_{-1}.
\label{f-equation_new}
\end{equation}
As  $f_{+1}+f_{-1}=1$,  one can easily obtain the solutions
\begin{equation}
f_{\pm}(t)=\frac{1-(1-2f_{\pm}(0))e^{-2pt}}{2},
\label{fsolve}
\end{equation}
where $f_{\pm1}(0)$  are the values of $f_{\pm 1}$  when $f_0$ reaches $0$. These equations are valid with the origin of time shifted  but it does not matter as we are interested in the $t \to \infty$ results. 
When $p=0$ we get the result that  $f_{\pm }(t \to \infty) \rightarrow f_{\pm }(0)$. This will then result in an ordered state (but not a consensus state in general).
Here it is assumed that the initial state is ordered,  for initial disordered states, $\langle O(t) \rangle=0$ for all times  as already mentioned.  
For any  nonzero value of  $p$ the system will reach an equilibrium state only with $\langle O\rangle =0$ as 
$f_{\pm 1} \to 1/2$ according to eq \ref{fsolve}. 
Thus for the $q=1$ point, we see that any $p \neq 0$ makes the system disordered.
 

For  values of $q \ne 1$, 
equation \ref{Oequation} indicates that 
in the steady state (at $t \rightarrow \infty $),  the system may reach an ordered state with  $\langle O\rangle \ne 0$ when $[-qp - p + f_0(1 - p + qp)]=0$, i.e., $f_0 = \frac{qp+p}{qp-p+1}$ (this puts a restriction $p\leq 0.5$ as $f_0 \leq 1$, hence no ordered state is possible  if $p \ge  0.5$). 
Now at the steady state  if we also demand that the individual densities 
 attain a fixed point, i.e.,  $\frac{df_{+1}}{dt}=0$ etc., then from equation  \ref{f+equation},
putting  $f_0 = \frac{qp+p}{qp-p+1}$, we get,

\begin{equation}
f_{+1}=\frac{-q-2p+2pq+1 \pm \sqrt{(1-q)(1-q-4p)}}{-2(p+q-2pq+pq^2 - 1)}.
\label{f+equation-2}
\end{equation}
Note that the above is valid for  $q \neq 1$. 

Therefore  the expression of  $\langle O\rangle$  in the steady state will be,

\begin{eqnarray}
\langle O\rangle&=&\frac{-q-2p+2pq+1 \pm \sqrt{(1-q)(1-q-4p)}}{-(p+q-2pq+pq^2 - 1)} \nonumber\\
& +&\frac{2p-1}{pq-p+1}.
\label{Oequation-2}
\end{eqnarray}
As $\langle O\rangle$ is  real,  a nonzero solution for $\langle O\rangle$ implies $(1-q-4p)$ should be greater than or equal to zero. This provides a  
a more stringent bound for the ordered phase given by  $\frac{1-q}{4}\geq p$.

\begin{figure}[h]
\centering
\includegraphics[width=7cm,angle=0]{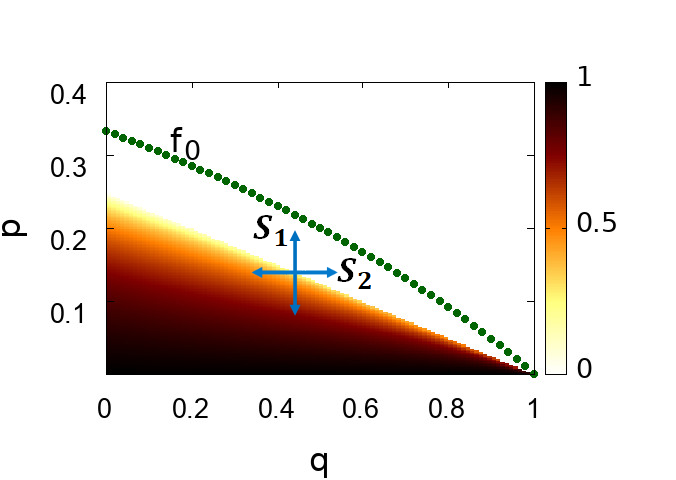}
\caption{Value of the order parameter in the $p-q$ phase space indicates the existence of the phase boundary given by $p=\frac{1-q}{4}$.
The dashed line in the disordered phase represents $f_0=\frac{1-q}{3-q}$.} 
\label{fig:phase_pic} 
\end{figure}

On the other hand, when the steady state (at $t \rightarrow \infty $) is   disordered,   $f_{+1}=f_{-1}$,  which when put in equation \ref{f+equation}, one gets 
\begin{gather}
    f_{+1}=f_{-1}=\dfrac{1}{3-q}; f_0=\dfrac{1-q}{3-q}.
\label{frozen-pt}    .
\end{gather}
Interestingly, the above values are independent of $p$.

\begin{figure}[h]
	\centering
	\includegraphics[width=9cm,angle=0]{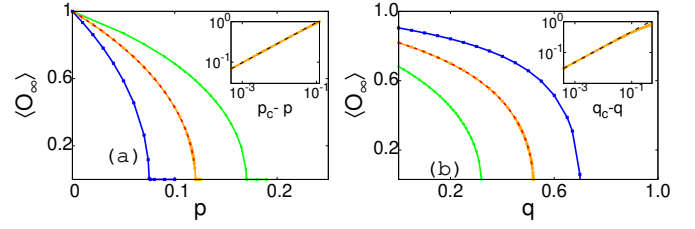}
	\caption{$\langle O_{\infty}\rangle$ calculated  numerically by solving eq. \ref{f+equation} and \ref{f-equation} for different sets of values of $q_c$ and $p_c$ along paths $S_1$ and $S_2$  respectively 
on the phase boundary 
(see Fig. \ref{fig:phase_pic}). 
In (a) the curves are for the path $S_1$ with fixed critical values of $q$; $q_c=0.70,0.52,0.32$ from left to right; in (b)  the curves are for the path $S_2$ with fixed critical values of $p$; $p_c=0.17,0.12,0.075$ from left to right. Analytical  values of  $\langle O\rangle$ from equation \ref{Oequation-2} are also plotted with dashed lines for  $q_c=0.52$ in (a) and $p_c=0.12$ in  (b).
Insets of (a) and (b) show the scaling behaviour of $\langle O\rangle$, indicating $\beta=0.5$} 
	\label{fig:O-p-q_pic} 
\end{figure}

{\it{Phase boundary:}} 

At the critical point between ordered-disordered phase transition  the fraction $f_0$ at the steady state for both the phases should be equal which gives
\begin{gather*}
    f_0=\dfrac{1-q}{3-q}= \frac{qp+p}{qp-p+1}.
\end{gather*}
Hence one gets an   equation of a straight line
\begin{equation}
p=\frac{1-q}{4}
\label{SAequation3}
\end{equation}
which is the phase boundary in the $p-q$ phase space shown in  Fig \ref{fig:phase_pic}. 
The value of $f_0$, a function of $q$ only in the disordered phase, is also shown.
Note that the order-disorder boundary obtained at $p=0$ for $q=1$ can be obtained as an analytical continuation from  the above equation.  However, all results discussed henceforth are for $q \ne 1$ in general.

{\it {Behaviour of  $\langle O\rangle$ close to a critical point}}:

Each point on the phase boundary is a critical point. We analyse the behaviour of the order
parameter close to  a critical point along  two different routes  $S_1$ and $S_2$ as indicated 
in Fig \ref{fig:phase_pic}. 
For path $S_1$, 
we take  $x=p_c - p$ and $q=q_c$ to  get from 
equation \ref{Oequation-2}, 
\begin{eqnarray}
\langle O\rangle&=&\frac{2(4x -4q_cx-q_c^2+1 \pm 4\sqrt{(1-q_c)x})}{(4x - q_c -8q_cx + 4q_c^2x-3q_c^2+q_c^3+3)} \nonumber\\
& -& \frac{\frac{q_c}{2} +2x+\frac{1}{2}}{\frac{q_c}{4} + x-q_c(\frac{q_c}{4} +x-\frac{1}{4})+\frac{3}{4} }.
\label{Oequation-3}
\end{eqnarray}
As $x \rightarrow 0$  behaviour of $\langle O\rangle \rightarrow \sqrt{x}$,  i.e., the critical exponent $\beta$ is $0.5$ along the path $S_1$.

Similarly, for path $S_2$, rewriting  equation \ref{Oequation-2} taking $y=q_c - q$ and $p=p_c$ we get
\begin{eqnarray}
\langle O\rangle &=&\frac{2p_cy - y - 4p_c +8p_c^2 \pm \sqrt{y(4p_c+y)}}{16p_c^3 + 8p_c^2y +p_cy^2 -4p_c -y} \nonumber \\
& -& \frac{2p_c-1}{p_c +p_c(4p_c+y-1)-1}.
\label{Oequation-4}
\end{eqnarray}
Thus  $\langle O\rangle \rightarrow \sqrt{y}$ as $y \rightarrow 0$ showing that  
 the value of the exponent $\beta=1/2$  does not depend on the path. In fact, the full variation of 
 $\langle O \rangle $ can easily be seen to depend on $[x+4y]^{1/2}$   as  the leading order term if we allow both $p$ and $q$ to vary about the critical point as before, i.e., $p=p_c -x$ and $q=q_c -y$ with the restriction that $x+4y \ge 0$ for a general direction. Note that for $S_1$ and $S_2$, both $x,y \ge 0$.

We have also numerically solved the time evolution equations to obtain the values of $\langle O(t \rightarrow \infty)\rangle$ along paths $S_1$ and $S_2$ for a particular point on the phase boundary to find that indeed 
the results are  compatible with $\beta = 0.5$ 
shown in  Fig \ref{fig:O-p-q_pic}.  

\subsection{Stability Analysis}

In the   disordered phase, we obtained a fixed point  characterised by   $f_{+1}=f_{-1}=\frac{1}{3-q}$. 
As there are three variables, in principle it is possible that in the disordered state, the values of $f_{+1}$ and $f_{-1}$  still evolve remaining the same.  
However, for the above  values  there can be no further change and hence   we call this the   
 frozen fixed point \cite{kb2}.

Stability of this frozen fixed point can be studied by introducing a deviation $\delta$ about it.   Taking  $f_{+1}= x^* + \delta$ and $f_{-1}=x^* - \delta$, 
where $x^*=\dfrac{1}{3-q}$, 
a  stability analysis leads to  $\delta(t)=\delta_0 \exp[\gamma t]$. Here $\delta_0$ is the initial deviation considered about the fixed point at $t=0$
and  
\begin{equation}
\gamma=\frac{1-q-4p}{3-q}.
\label{SAequation_coeff}
\end{equation}
As $\langle O\rangle=f_{+1}-f_{-1}=2\delta$,  we get
\begin{equation}
\langle O\rangle=2\delta_0 \exp[\gamma t].
\label{SAequation2}
\end{equation}
The above equation shows that for an initially ordered state, there will be a growth/decay of the 
order parameter according to the sign of $\gamma$ which changes at $1-q-4p =0$. This is consistent with 
the phase boundary at $1-q= 4p$ and  as expected one gets a disordered state even when starting from an ordered 
state in the disordered phase. 

\begin{figure}[h]
	\centering
	\includegraphics[width=9cm,angle=0]{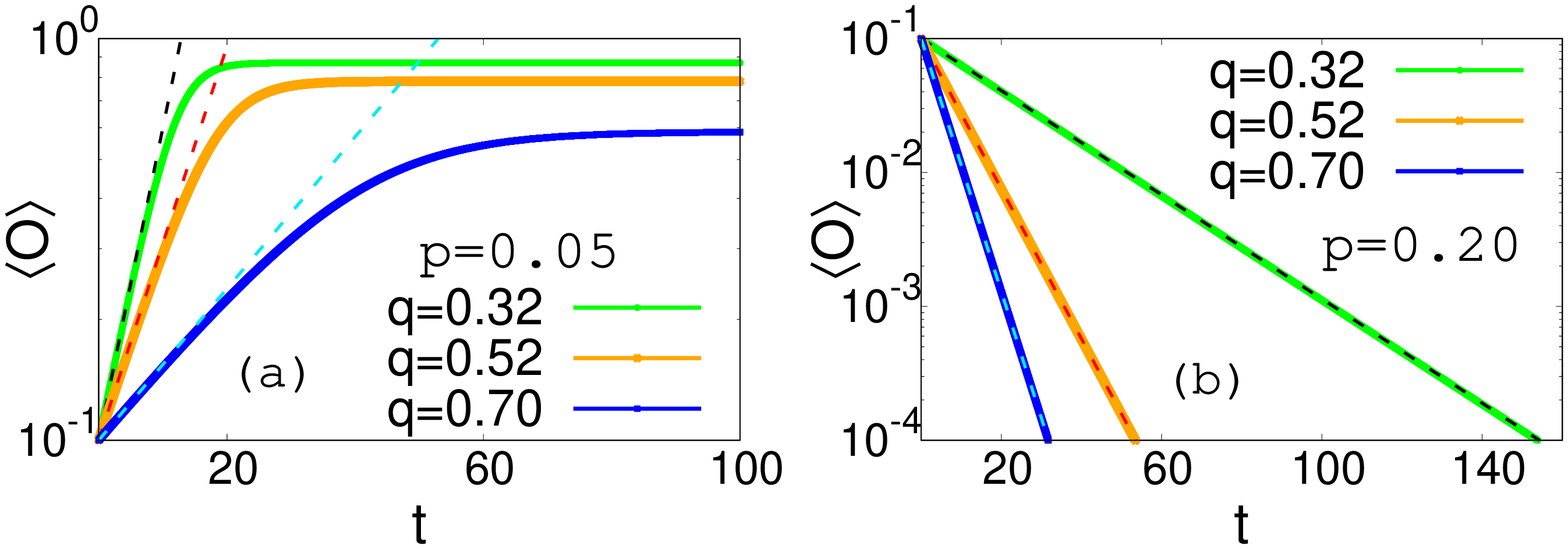}
	\caption{The time evolution of $\langle O\rangle$ generated numerically from eq. \ref{f+equation} and \ref{f-equation} for $\frac{1-q}{4}>p$ (ordered phase) and $\frac{1-q}{4}<p$ (disordered phase) are shown in (a) and (b) respectively with the initial configuration  $f_{+1}=x^* + \delta_0$ and $f_{-1}=x^* - \delta_0$ (see text). The dotted line in each graph represents the fitted graph following  equation \ref{SAequation2}.  } 
	\label{fig:O-delta_pic} 
\end{figure}

Equation \ref{SAequation2}  shows that there is a time scale $\tau = \gamma^{-1}$ associated with
the dynamics of the growth/decay and  $\tau$  diverges at the phase boundary indicating critical
slowing down. Once again, the time dependent equations are solved numerically  by taking initial states close to the frozen fixed point and the results agree with the above as shown in Fig \ref{fig:O-delta_pic}.    Since $\tau \propto (1-q-4p)^{-1}$, it   diverges  with an exponent   
 $1$ which is related to the critical dynamical exponent $z$, this is to be discussed 
further in the next section.

%

\subsection{Relaxation from a perfectly ordered state}

While for initial states with small order, the order parameter  will either decay or grow depending on whether one is
in the disordered or ordered phase, for a fully ordered initial state  the order parameter will
decrease in time  
 in the in both phases. 
We study the relaxation behaviour by numerically solving the rate equations taking initial condition $f_{+1} = 1$.

\begin{figure}[h]
	\centering
	\includegraphics[width=9cm,angle=0]{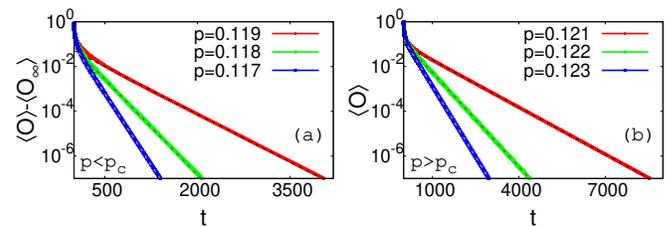}
	\caption{Time evolution of $\langle O\rangle$ calculated by  numerically solving  eq. \ref{f+equation} and \ref{f-equation} with a fixed value of  $q=0.52$ and  different values of $p$ (a) below and (b) above the corresponding $p_c$ when the initial condition is $(f_{+},f_0,f_{-1}) =(1,0,0)$.
 The dotted lines show the fitted curve with equation \ref{TSequation1}.} 
	\label{fig:O-p-q-tau-1-pic} 
\end{figure}

\begin{figure}[h]
	\centering
	\includegraphics[width=9cm,angle=0]{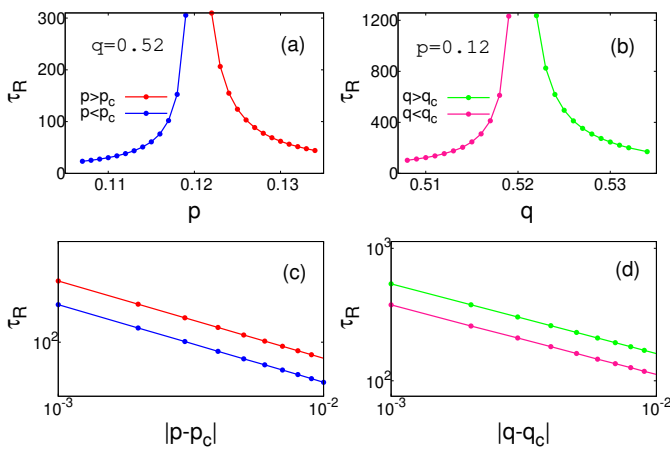}
	\caption{The time scale $\tau_R$ calculated from the slopes of the log-linear plots in Fig \ref{fig:O-p-q-tau-1-pic} using 
equation \ref{TSequation1} is shown  
(a)  against  $p$ for a  constant value of $q$ and (b) against   $q$ for a constant value of $p$. In (c) and (d),  these values of $\tau_R$  are plotted against $|p-p_c|$ and $|q-q_c|$ respectively, both above and below the critical values to show a power law divergence  with the 
associated exponent very close to unity. The color codes for the upper and lower panels are same.} 
	\label{fig:O-p-q-tau-2-pic} 
\end{figure}

In the disordered phase the decay of the order parameter  is expected to  
follow 
Eq \ref{SAequation2}  at late times as it comes closer to the frozen fixed point, indicating again the presence of a time scale $\propto |\gamma|^{-1}$. 
In the ordered phase, the order parameter will initially decay and then attain a nonzero saturation 
value. We find that both behaviour are captured by a single equation
\begin{equation}
(\langle O\rangle-\langle O_{\infty}\rangle) \propto \exp[-t/\tau_R] 
\label{TSequation1}
\end{equation}
where $\langle O_{\infty}\rangle$ is the ensemble averaged equilibrium value of the ordered parameter at $t\rightarrow \infty$.
We have plotted the   data for some  particular points above and below the phase boundary in  Fig \ref{fig:O-p-q-tau-1-pic},
and the timescales $\tau_R$ extracted from the slopes of the log-linear graphs are shown in Fig. \ref{fig:O-p-q-tau-2-pic}a,b.  
The results  show that   $\tau$ and $\tau_R$ have identical scaling behaviour 
in the disordered phase as argued above, while 
in the ordered phase also,
$\tau_R$   diverges close to the critical point with the same exponent 1
(see Fig \ref{fig:O-p-q-tau-2-pic}c,d)

Lastly, we plot $\langle O\rangle$ as a function of time at several points exactly on the phase boundary in Fig. \ref{fig:O-t-critical-pic} to get a power law decay with an exponent close to 0.5.  
This discussion of course excludes the  $q = 1$ point which is unique, one can easily check   that with the critical value $p=0$ here, no evolution of the   initially 
fully ordered  state is possible.



\textsc{\begin{figure}[h]
	\centering
	\includegraphics[width=7cm,angle=0]{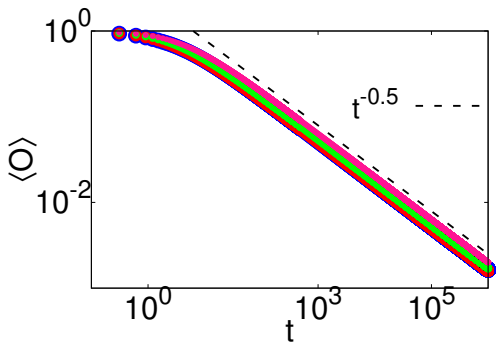}
	\caption{Time variation of $\langle O\rangle$ calculated  by numerically solving eq. \ref{f+equation} and \ref{f-equation} exactly at  different critical points on the phase boundary indicates   $O(t\rightarrow \infty)\propto t^{-0.5}$
.} 
	\label{fig:O-t-critical-pic} 
\end{figure}}

%



\section{Summary and Discussions}

In this paper, we have studied the case of extreme switches in opinion in a three-state kinetic exchange model where the interactions may be both positive as well as negative. The two parameters characterising the probabilities of the extreme switches and 
negative interactions are $q$ and $p$ respectively. Our main findings are\\
$\bullet$ The presence of a phase boundary given by $p=\frac{1-q}{4}$\\
$\bullet$ Exponent $\beta$ associated with the order parameter is universal with the  value $1/2$\\
$\bullet$ The phase boundary can also be obtained using stability analysis.
Additionally one gets   the  time evolution of the  partially ordered state showing 
  exponential growth/decay. The associated time scale $\tau$ diverges with an exponent $-1$.  \\
$\bullet$ Relaxation behaviour of the fully ordered state shows the expected exponential decay of the order parameter with time in the disordered phase; for the ordered phase, it relaxes exponentially to a saturation value. The relaxation timescale $\tau_R$ and $\tau$ have identical scaling behaviour. Exactly on the phase boundary, the order parameter shows a power law decay. \\
$\bullet$ The overall behaviour is mean field  like for $q \neq 1$. \\
$\bullet$ The $q=1,p=0$ has a special significance. 

\begin{figure}[h]
	\centering
	\includegraphics[width=7cm,angle=0]{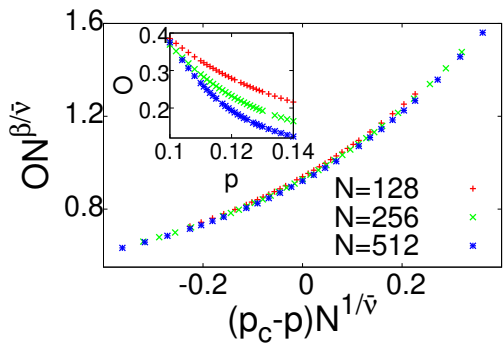}
	\caption{Simulation results: Data collapse  
of the scaled order parameter  $ON^{\beta/\bar \nu}$, when plotted against  $(p_c-p) N^{1/\bar \nu}$ for a fixed value of $q=0.52$, is obtained
 using the values   $\bar{\nu}=2.0$ and $\beta=0.5$. Inset shows the raw data.} 
	\label{fig:simulation1} 
\end{figure}

While the first four issues have already been discussed in detail in the previous section, we focus on the last two points here. 
It had been already  known that in the mean field three state kinetic exchange model without extreme switches, the value of the exponent $\beta =1/2$.
Also,    assuming the mean field model has an effective dimension $d$, the exponent $\bar \nu = \nu d = 2$ was obtained previously where $\nu$ is the correlation length exponent 
 \cite{BCS}.
We have found $\beta$ to be equal to 1/2 at any point   on the phase boundary in the 
two parameter model for $q \ne 1$.
To get $\bar \nu$, we conduct  small scale simulations about a 
particular point on the phase boundary.  
Indeed the scaled order parameter  
curves collapse when plotted against $|\epsilon| N^{1/{\bar \nu}}$, where $\epsilon$ denotes the deviation from the critical point,  
with  $\beta=1/2$ and $\bar \nu =2$. 
The raw data and the collapse are shown in Fig \ref{fig:simulation1}.   
Hence the static critical behaviour is unaffected by the parameter $q \ne 1$. 

The critical slowing down phenomena is observed with a timescale diverging as $|\epsilon|^{-1}$, again independent of $q$.
Since in a continuous phase transition, the time scale diverges  as $\xi^{z}$ where $\xi$ is the 
correlation length $\propto |\epsilon|^ {-\nu}$ and $z$ the dynamic critical exponent,  one gets   
$\tau \propto |\epsilon|^{-\nu z}$. Hence our results indicate  $\nu z =1$. 

Next one can consider the relaxation of the order parameter exactly at criticality.  The power law 
behaviour $t^{-1/2}$ can be shown to be compatible with 
the theory of dynamic critical phenomena.
In general  the dynamical behaviour of the order parameter is given by 
 $O(t) \propto t^{-y} f(t/\xi^z)$ with $O(t) \propto t^{-y}$ exactly at the critical point. 
Equilibrium behaviour indicates $y =  \beta/\nu z$. Therefore, 
 using  the values  $\beta=1/2$ and $\nu z = 1$, 
 one gets $y=1/2$, which is the   value obtained  here. 
We also remark that the values of the static and dynamic exponents 
obtained here coincide with those of the 
 mean field Ising model ($\beta = 1/2, \nu = 1/2, z=2$) if one uses $d=4$, the upper critical dimension of the Ising model.

Even though the critical behaviour is unchanged with a nonzero value of $q$, 
we note that the fixed point values are independent of $p$ in the disordered phase. Also, the variation of $f_0$ in the disordered phase (see Fig. \ref{fig:phase_pic})  suggests that as $q$ is increased,  the model
tends to become a binary one. 
The interpretation  is, with increasing probability of extreme switches, 
the system in the disordered phase  tends to be polarised as it becomes 
 increasingly difficult to retain a neutral opinion. 

For $q=1$,  we get  two equations (\ref{f+equation_new} and \ref{f-equation_new}) after a transient time when $f_0$ becomes zero.  
These can be easily identified  as  
the  equations governing the  dynamics of a two state  voter model with  random flipping probability $p$.  Obviously, it becomes disordered at any 
value of $p$. 
Previously it was noted that for $p=0$, the
model is identical to the mean field voter model for $q=1$, 
with $p \ne 0$ we thus obtain a 
mapping to a voter model with random flipping.  

Hence the main conclusion from the present study is that the extreme switches act as additional noise for the model considered in \cite{BCS} thereby lowering the critical values $p_c$ without 
changing the critical behaviour. The point $q=1$ has a special interpretation. 
The role of the two kinds of disorder are however different, the system 
becomes disordered for a finite value of $p$ (for $q=0$) but remains ordered  up to the extreme value of $q$ (for $p=0$) \cite{kb2}.
It is, therefore, not surprising that the critical behaviour is dominated by $p$ while $q$  acts as an irrelevant variable.
However, the nature of the disordered phase is dictated by $q$ alone.

The results obtained in the present paper are based on the mean field dynamical equations. Of course,  if we consider the system on lattices with nearest neighbour coupling, there will be quantitative changes. 
As a future study, it will be interesting to investigate how the 
 extreme switches affect qualitatively the criticality and dynamics in finite dimensions.

{\bf {Acknowledgement}}

PS acknowledges financial support from SERB (Government of India) through scheme no MTR/2020/000356. We thank  Sudip Mukherjee, Soumyajyoti Biswas and Arnab Chatterjee   for some discussions.

\end{document}